\documentclass[aps,twocolumn,showpacs,showkeys,amsmath]{revtex4}
\usepackage{epsfig,graphicx}
\usepackage{dcolumn}
\usepackage{bm}

\begin{document}

%\begin{Large}

\title{Theoretical Evaluation of the Reaction Rates for $^{26}$Al$(n,p)^{26}$Mg and $^{26}$Al$(n,\alpha)^{23}$Na}

\author{B. M. Oginni,\footnote{oginni@physics.unc.edu} C. Iliadis and A. E. Champagne}
\address{Department of Physics and Astronomy, University of North Carolina at Chapel Hill, NC 27599 and \\Triangle Universities Nuclear Laboratory, Durham, NC 27708}

\begin{abstract}
The reactions that destroy $^{26}$Al in massive stars have significance in a number of astrophysical contexts. We evaluate the reaction rates of $^{26}$Al$(n,p)^{26}$Mg and $^{26}$Al$(n,\alpha)^{23}$Na using cross sections obtained from the codes EMPIRE and TALYS. These have been compared to the published rates obtained from the NON-SMOKER code and to some experimental data. We show that the results obtained from EMPIRE and TALYS are comparable to those from NON-SMOKER. We also show how the theoretical results vary with respect to changes in the input parameters. Finally, we present recommended rates for these reactions using the available experimental data and our new theoretical results.
\end{abstract}

\maketitle
%\textbf{Keywords: } Level Density

\section {INTRODUCTION} \label{sec0}

$^{26}$Al $\beta$-decays to the first excited state of $^{26}$Mg,  which emits a $\gamma$-ray of 1809 keV energy. These $\gamma$-rays were initially detected by the third High Energy Astronomy Observatory (HEAO3) satellite \cite{heao}, the Solar Maximum Mission Satellite \cite{solarmax}, Gamma-Ray Imaging Spectrometer (GRIS) \cite{gris}, and others in the last century. Subsequently, an all-sky image of $^{26}$Al $\gamma$-ray emission at 1809 keV was derived from the COMPTEL instrument on-board the Compton Gamma Ray Observatory (CGRO) \cite{pluschke}, and further refined by INTEGRAL data, which showed that the $^{26}$Al co-rotates with the galactic disk \cite{diehl}. This map provides information about star formation and nucleosynthesis over a time period comparable to the half-life of $^{26}$Al (7.17 $\times$ 10$^{5}$ y).

 It has also been observed that $^{26}$Mg is overabundant in some  meteoritic inclusions, which is attributed to the decay of $^{26}$Al \cite{leet}. It implies that live $^{26}$Al was present during the formation of the meteorites. Since the parent bodies of meteorites were formed during the early stages of the solar system, this observation may provide information about the last nucleosynthesis events that contributed matter to the solar nebula. The Galactic abundance of $^{26}$Al depends largely on the rates of the reactions that lead to its production and destruction. The main production mechanism in a variety of sites, such as massive stars or AGB stars, is the $^{25}$Mg$(p,\gamma)^{26}$Al reaction, while the main destruction mechanisms at higher temperatures are the $^{26}$Al$(n,p)^{26}$Mg and $^{26}$Al$(n,\alpha)^{23}$Na reactions \cite{woosleywe, limongich}.

Massive stars may produce $^{26}$Al during different phases of their evolution: (i) during pre-supernova stages in the C/Ne convective shell, near temperatures of $\approx$ 1 GK, where a fraction of the $^{26}$Al survives the subsequent explosion and is ejected into the interstellar medium \cite{arnett78}; (ii) during core collapse via explosive Ne/C burning \cite{arnett77}, near temperatures of $\approx$ 2 GK, where the ejected $^{26}$Al yield may perharps be modified by the $\nu$-process via neutrino spallation \cite{woosley90}; and (iii) in Wolf-Rayet stars, i.e., stars with masses in excess of about 30 \textit{M$_{\odot}$}, during core hydrogen burning, near temperatures of $\approx$ 0.1 GK, where $^{26}$Al may appear via convection at the surface and, subsequently, is ejected by strong stellar winds \cite{palacios05}. These $^{26}$Al production mechanisms (with the exception of the $\nu$-process) were recently analyzed in detail by Limongi and Chieffi \cite{limongich} by using extensive hydrodynamic simulations of solar metallicity stars in the mass range of 11\textit{M$_{\odot}$} $\le M \le$ 120\textit{M$_{\odot}$}. In that work, they also emphasized the impact of rate uncertainties for selected reactions on the final $^{26}$Al yields.

Direct measurement of the important $^{26}$Al$(n,p)^{26}$Mg and $^{26}$Al$(n,\alpha)^{23}$Na reactions are challenging. Thus the cross sections of these reactions are evaluated in this work with the codes EMPIRE \cite{empire} and TALYS \cite{talys} using different prescriptions for level densities and optical model potentials. In Sec. \ref{sec1}, we describe the experimental status of these reactions. In Sec. \ref{sec2}, we discuss the underlying nuclear reaction theory. In Sec. \ref{sec3}, we examine the effects of variations in the level density and the optical model parameters, and in Sec. \ref{sec4}, we present recommended reaction rates. Reaction rates involving the isomeric state in $^{26}$Al are discussed in the Appendix.

%in Sec. \ref{sec2}, we discuss the effects of variations in the models used for the level density; in Sec. \ref{sec3}, we examine how the results vary with changes in the optical-model parameters and in Sec. \ref{sec4}, we compare all the theoretical results and draw some conclusions.

\section{EXPERIMENTAL STATUS} \label{sec1}
Despite the astrophysical importance of the reactions that destroy $^{26}$Al, the available data are sparse. This is largely because $^{26}$Al is radioactive and its natural abundance is very small. Thus the preparation of a suitable target becomes a challenge. In order to overcome this hurdle, earlier experimental efforts focussed on measuring the inverse reactions $^{26}$Mg$(p,n)^{26}$Al and $^{23}$Na($\alpha,n)^{26}$Al and, using the principle of detailed balance,  determined the forward reaction cross sections for the transitions to the ground states of $^{26}$Mg or $^{23}$Na. Unfortunately, this procedure does not provide any information on reaction channels populating excited states, which may represent the dominant contributions. Thus it is clearly of advantage to perform direct measurements using a radioactive $^{26}$Al target. The only experimental data available are those of Refs. \cite{smet, koehler, traut, skelton}. 

\begin{figure}[!htb]
\begin{center}
\includegraphics[angle=270, totalheight=0.35\textheight, width=0.5\textwidth]{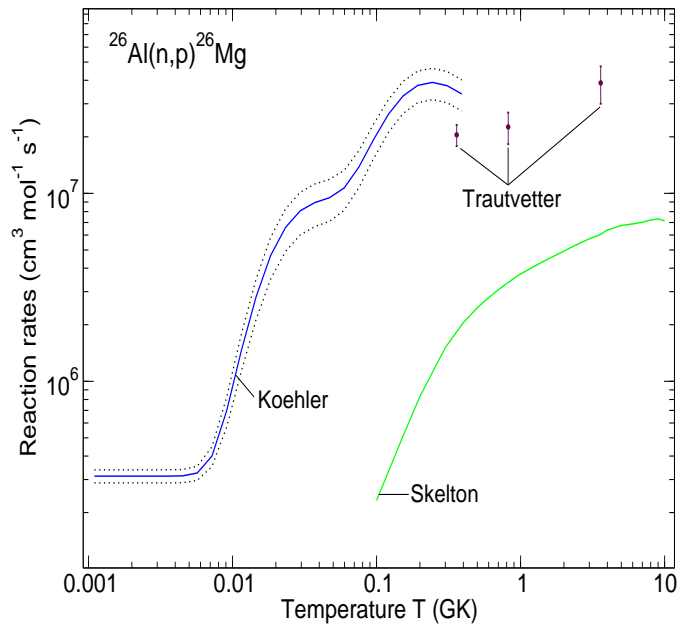}
\caption{ (Color online): The experimental reaction rates for $^{26}$Al$(n,p)^{26}$Mg, as obtained by Koehler \textit{et al.} \cite{koehler} in blue, with the uncertainties represented by dotted lines; Trautvetter \textit{et al.} \cite{traut} (data points); and Skelton \textit{et al.} \cite{skelton} in green. The latter results are deduced from measurement of the inverse reaction and represent the contribution of the ($n,p_{0}$) channel only.}
\label{exp1}
\end{center}
\end{figure}

\begin{figure}[!htb]
\begin{center}
\includegraphics[angle=270, totalheight=0.35\textheight, width=0.5\textwidth]{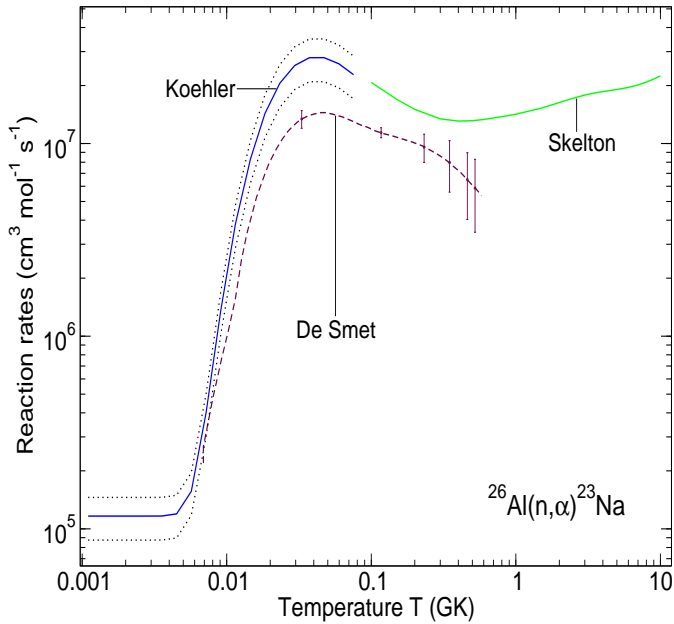}
\caption{ (Color online): The experimental reaction rates for $^{26}$Al$(n,\alpha)^{23}$Na as obtained by Koehler \textit{et al.} \cite{koehler} in blue, with the uncertainties represented by dotted lines; De Smet \textit{et al.} \cite{smet} represented by the dashed lines where the vertical lines indicate the uncertainty range; and  Skelton \textit{et al.} \cite{skelton} in green. The latter rates are deduced from measurement of the inverse reaction and represents the contribution of the ($n,\alpha_{0}$) channel only.}
\label{exp2}
\end{center}
\end{figure}

Figure \ref{exp1} shows the experimental reaction rates for $^{26}$Al$(n,p)^{26}$Mg. The rates of Skelton \textit{et al.} \cite{skelton} were based on the inverse reaction, while the rates from Koehler \textit{et al.} \cite{koehler} are for the  ($n,p_{1}$) channel, which is believed to be the dominant contribution. The rates from Trautvetter \textit{et al.} \cite{traut} combine results for the $p_{0}$ and $p_{1}$ transitions. The results from Skelton \textit{et al.} \cite{skelton}  represent only a fraction of the total rates. It is evident from the figure that the Koehler \textit{et al.} \cite{koehler} and Trautvetter \textit{et al.} \cite{traut} results are not consistent. Figure \ref{exp2} shows the rates for $^{26}$Al($n,\alpha$)$^{23}$Na. The Skelton \textit{et al.} \cite{skelton} rates are deduced from the inverse reaction. Koehler \textit{et al.} \cite{koehler} only measured the ($n,\alpha_{0}$) channel, which is the dominant contribution. The results of De Smet \textit{et al.} \cite{smet} represent the total contribution from $^{26}$Al($n,\alpha_{0} + \alpha_{1}$)$^{23}$Na. Again, the available data are not consistent. It should be noted that the data from Koehler \textit{et al.} \cite{koehler} and De Smet \textit{et al.} \cite{smet} provide rates at temperatures much lower than what is needed to model $^{26}$Al synthesis in massive stars (Sec. \ref{sec0}). Clearly, there is a need for more data. However, in the absence of new experimental data, we
resort for now to theoretical calculations and investigate how these compare with the current data.

\section{THEORY} \label{sec2}

Nuclear reactions may have contributions from direct, pre-equilibrum and compound mechanisms, depending on the time taken for the reaction to occur with respect to the time it takes for the projectile to traverse the target nucleus. In this work, we
calculate the cross sections of $^{26}$Al$(n,p)^{26}$Mg and $^{26}$Al$(n,\alpha)^{23}$Na using the nuclear reaction codes  EMPIRE \cite{empire} and TALYS \cite{talys}.  

%The NON-SMOKER code is popular in nuclear astrophysics, but in this study we have given attention to the TALYS and EMPIRE codes, which are widely used in other fields of nuclear physics, for a number of reasons. NON-SMOKER code takes into account only the compound reaction mechanism using the Hauser Feshbach model. The TALYS code takes into account all possible reaction mechanisms, while EMPIRE has the option to include other reaction mechanisms besides compound reactions. In addition, users of TALYS and EMPIRE can vary the different input parameters they desire. However, in NON-SMOKER, it is possible to change the input parameters but the code is not in distribution, so users have no control over the input.

The codes  use different nuclear models to calculate the contributions from the various reaction mechanisms. The EMPIRE code estimates the direct reaction contribution using the distorted wave Born approximation (DWBA) and coupled-channels calculations according to  Refs. \cite{raynal, dasso}; the preequilibrum reaction contribution from Refs. \cite{feshbach, nishioka, tamura, blann}; and the compound reaction contribution from Hauser Feshbach theory \cite{hauser}, with the width fluctuation correction implemented in terms of the HRTW (according to the names of the authors, Hofmann, Richert, Tepel, Weidenm$\ddot{u}$ller) approach \cite{hrtw1, hrtw2}. The TALYS code computes the direct reaction contribution using the DWBA for (nearly) spherical nuclei; the coupled-channels model for deformed nuclei from Ref. \cite{tamura2} and the weak-coupling model for odd nuclei from Ref. \cite{hodgson}; the preequilibrum reaction contributions from exciton models \cite{koning2, gruppelaar, gadioli2, akkermans} and Kalbach systematics \cite{kalbach}; and the compound reaction employs the Hauser Feshbach model \cite{hauser}, including width fluctuation corrections \cite{hofmann, tepel, hofmann2, moldauer, moldauer2, verbaarschot}. 

%\clearpage

\begin{figure}[!htb]
\begin{center}
\includegraphics[angle=270, totalheight=0.35\textheight, width=0.5\textwidth]{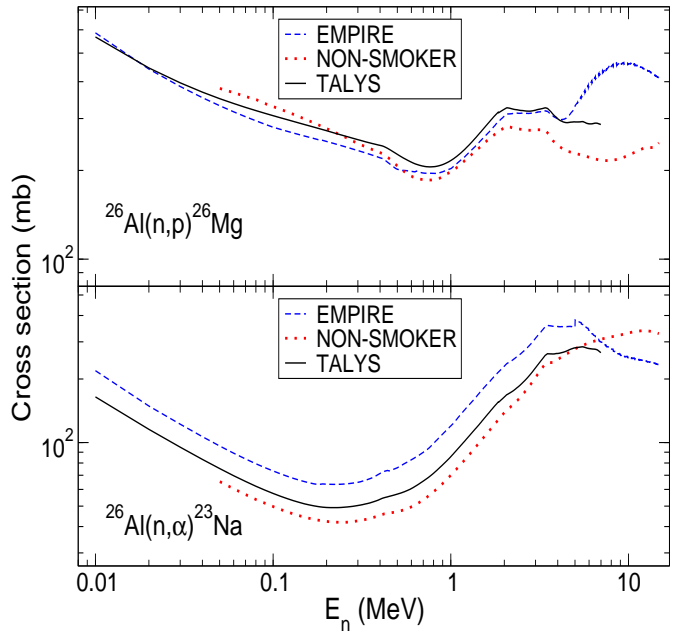}
\caption{ (Color online): The $^{26}$Al$(n,p)^{26}$Mg and $^{26}$Al$(n,\alpha)^{23}$Na cross sections as calculated by  EMPIRE \cite{empire}, TALYS \cite{talys} and NON-SMOKER \cite{rauscher1, rauscher2}.}
\label{thr1}
\end{center}
\end{figure}

Figure \ref{thr1} shows the cross section versus bombarding energy. The results from TALYS and EMPIRE are presented alongside the published NON-SMOKER \cite{rauscher1, rauscher2} calculations. The level density model options used in the TALYS and EMPIRE calculations are the constant temperature plus Fermi gas and the EMPIRE-specific models respectively, while the default option was used for the optical model potential parameters. Details of these are discussed in Sec. \ref{sec3}. The three theoretical results are largely in agreement to within a factor of 2. The cross sections calculated from the nuclear reaction codes can be used to evaluate the reaction rates via \cite{christian}:
% Numerically, we obtain the reaction rate at a given temperature, 

\begin{eqnarray}\label{eqn3}
N_{A}<\sigma v>_{01}& = &\frac{3.7318 \times 10^{10}}{T_{9}^{3/2}}\sqrt{\frac{M_{0} + M_{1}}{M_{0}M_{1}}}\nonumber\\
&&\times \int_{0}^{\infty} E\sigma(E)e^{-11.605E/T_{9}}dE\nonumber\\
 &&\textrm{(cm$^{3}$ mol$^{-1}$ s$^{-1}$)}
\end{eqnarray}
where $E$ is the center-of-mass energy in units of MeV, $T_{9}$ is the temperature in GK ($T_{9} \equiv T/10^{9}$K), $M_{i}$ are the relative atomic masses in u, and $\sigma$ is the cross section in barn (1 b $\equiv 10^{-24}$ cm$^{2}$).

\begin{figure}[!htb]
\begin{center}
\includegraphics[angle=270, totalheight=0.43\textheight, width=0.5\textwidth]{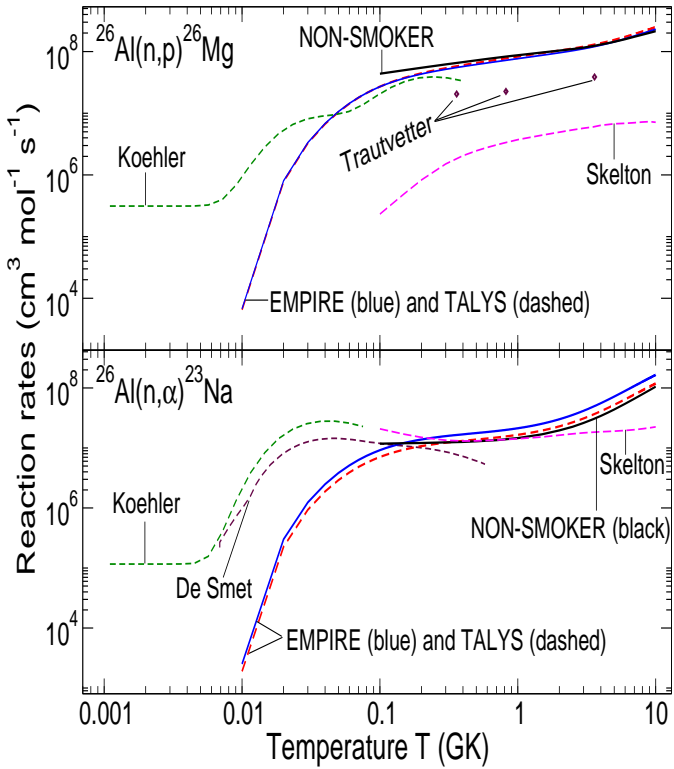}
\caption{ (Color online): The astrophysical reaction rates for $^{26}$Al$(n,p)^{26}$Mg (top) and $^{26}$Al$(n,\alpha)^{23}$Na (bottom) as obtained by different experiments and theoretical calculations.}
\label{thex1}
\end{center}
\end{figure}

Figure \ref{thex1} shows how the reaction rates from the theoretical calculations compare with the available experimental data. The theoretical calculations are based on statistical models, which are valid when the nuclei have sufficiently high level densities. This is the reason for the significant underprediction of the theoretical rates in the low temperature region. For $^{26}$Al$(n,p)^{26}$Mg, the theoretical calculations  are consistent with one another above 100 MK to within a factor of 1.7. They also agree with the results of Koehler \textit{et al.} \cite{koehler} at temperatures between T = 0.04 - 0.4 GK within a factor of 2. However, our theoretical results are higher than those of Trautvetter \textit{et al.} \cite{traut} by more than a factor of 2.5 on the average. As already noted, the Skelton \textit{et al.} \cite{skelton} rates are too low because they do not take into account the main ($n,p_{1}$) channel. For $^{26}$Al$(n,\alpha)^{23}$Na, the theoretical results are again in good agreement above 100 MK within a factor of 1.7. They agree with the experimental rate of De Smet \textit{et al.} \cite{smet} at T $\approx$ 0.1 - 0.3 GK. The comprism with the Koehler \textit{et al.}  \cite{koehler} rate is not so clear, since the latter rate cuts off at a maximum temperature of 80 MK. Below this temperature, the theoretical rate is not expected to be reliable because of the low level densities. The theoretical rates are in good agreement with Skelton \textit{et al.} \cite{skelton} at T = 0.2 - 2 GK, within a factor of 2. However at higher temperature, our rates become significantly larger since transitions to excited states become important, which are not taken into account by Ref. \cite{skelton}.

%The TALYS and EMPIRE codes also have some flexibilities in terms of the input parameters that are used, one can change the level density and optical parameters models used in the calculations; NON-SMOKER does not have such capability.

\begin{figure}[!htb]
\begin{center}
\includegraphics[angle=270, totalheight=0.3\textheight, width=0.5\textwidth]{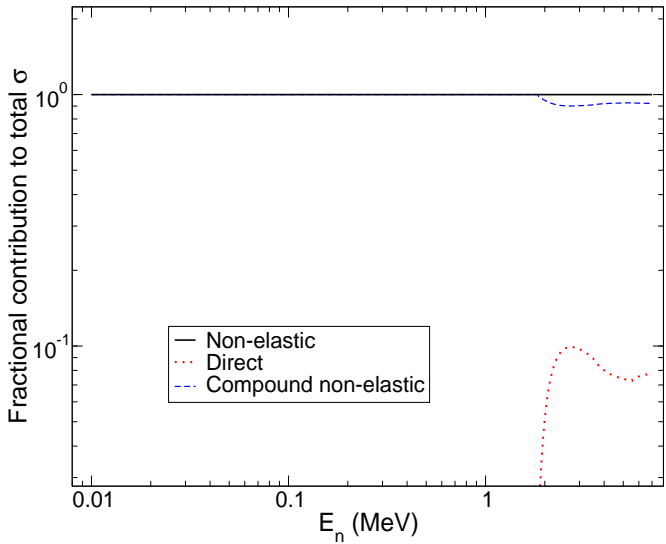}
\caption{ (Color online): Contributions from different reaction mechanisms to the total non-elastic cross section, as predicted by the TALYS code.}
\label{thr2}
\end{center}
\end{figure}

The contributions of the various reaction mechanisms to the total non-elastic cross section, calculated using TALYS, is shown in Fig. \ref{thr2}. Up to about 2 MeV bombarding energy, the main contribution to the non-elastic cross section arises from compound nuclear reactions. Since this energy region is the most relevant to the reaction rates, we can conclude that pre-equilibrum and direct processes are negligible. This explains the general level of agreement between the three codes. Nevertheless, it is interesting to see that there are differences, which are probably attributable to differences in the input parameters used in each calculation.

\section{VARIATION OF MODEL PARAMETERS} \label{sec3}

A major goal of the present study was to explore the sensitivity of the theoretical reaction rates to variations in the model parameters. The main ingredients of the Hauser-Feshbach (HF) model are the optical model parameters (used to determine the transmission coefficients) and the nuclear level densities. The latter are experimentally known up to some excitation energy beyond which one needs to employ a level density model. Thus most nuclear level density models agree at low excitation energies, but may differ greatly at higher excitation energies. 

%In this study, we used various optical model parameters and level density models provided in the codes to see how the cross sections and reaction rates deviate from the default option.

\begin{figure}[!htb]
\begin{center}
\includegraphics[angle=270, totalheight=0.35\textheight, width=0.5\textwidth]{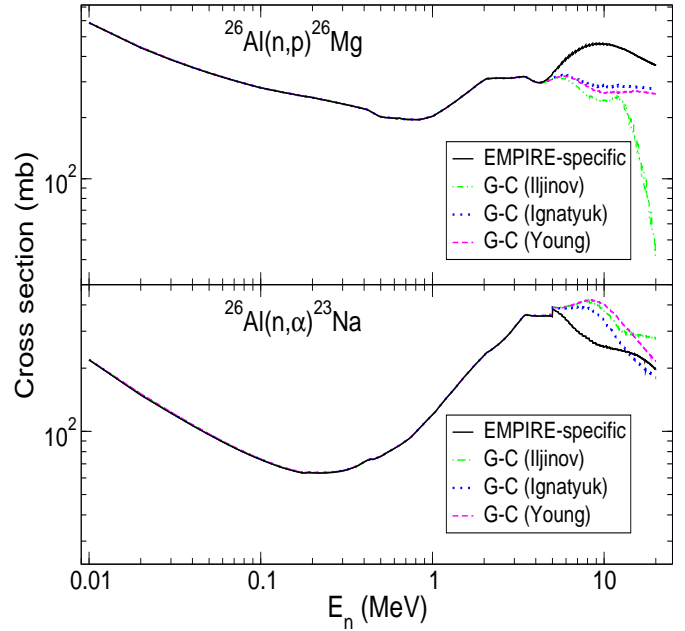}
\caption{ (Color online): The $^{26}$Al($n,p)^{26}$Mg (top) and $^{26}$Al$(n,\alpha)^{23}$Na (bottom) cross sections calculated using the EMPIRE code assuming different level density models; ``G-C'' stands for the Gilbert Cameron model described in the text.}
\label{fig1}
\end{center}
\end{figure}

We started by varying the level densities. For the EMPIRE code, we explored the EMPIRE-specific and the Gilbert Cameron (GC) models. The EMPIRE-specific model combines the super-fluid \cite{bcs} and Fermi-Gas (FG) models \cite{bethe} with deformation-dependent collective effects. The Gilbert Cameron model is a combination of the constant temperature (CT) model \cite{gilbert1, gilbert2} at low energy and the Fermi-Gas model \cite{bethe} at higher energy. This model has been parameterized by Ignatyuk \cite{bcs}, Iljinov \cite{iljinov},  and Young \cite{young}. Figure $\ref{fig1}$  shows the results of cross section calculations using EMPIRE with the different options for the level density model.  One can see that the cross sections obtained from these calculations are similar up to a certain value at high energy beyond which they appear to deviate from one another. This is not unexpected since the energy levels are experimentally known up to a particular excitation energy (i.e., the discrete region), while different theoretical prescriptions are used for higher energies (i.e., the continuum region).  The cross sections obtained were used to compute the reaction rates using Eq. (\ref{eqn3}). The rates are very similar in magnitude at astrophysically relevant temperatures (T = 0.01 - 10 GK). This is because the cross sections at low bombarding energies contribute the most to the rates, corresponding to low excitation energies at which the level densities are experimentally known.

%\begin{figure}[!htb]
%\begin{center}
%\includegraphics[angle=270, totalheight=0.35\textheight, width=0.5\textwidth]{graph_erld.ps}
%\caption{ (Color online): Reaction rates for $^{26}$Al$(n,p)^{26}$Mg and $^{26}$Al$(n,\alpha)^{23}$Na calculated using the EMPIRE code with different level density models.}
%\label{fig2}
%\end{center}
%\end{figure}

\begin{figure}[!htb]
\begin{center}
\includegraphics[angle=270, totalheight=0.35\textheight, width=0.5\textwidth]{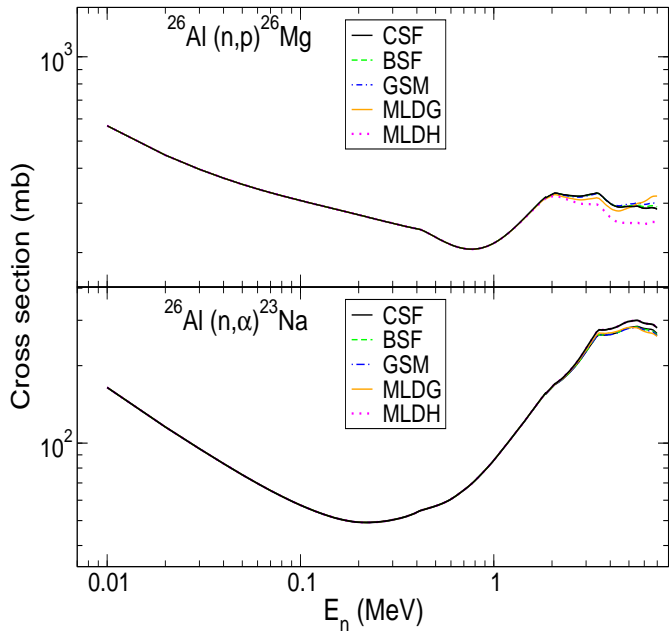}
\caption{ (Color online): The $^{26}$Al$(n,p)^{26}$Mg and $^{26}$Al$(n,\alpha)^{23}$Na cross sections as calculated using the TALYS code with different level density models. The legends are as follow: constant temperature plus Fermi gas model (CSF) \cite{gilbert1, gilbert2}; back-shifted Fermi gas model (BSF) \cite{bsf}; Generalized superfluid model (GSM) \cite{bcs, gsm}; Microscopic level densities from Goriely's table (MLDG) \cite{mldg}; and microscopic level densities from Hilaire's table (MLDH) \cite{mldh}.}
\label{figt1}
\end{center}
\end{figure}

For the TALYS code, we explored five different level density models: the constant temperature plus Fermi gas model (CSF), which is equivalent to the Gilbert-Cameron model \cite{gilbert1, gilbert2}; the back-shifted Fermi gas model (BSF) \cite{bsf}; the generalized superfluid model (GSM) \cite{bcs, gsm}; the microscopic level densities from Goriely's table (MLDG) \cite{mldg}; and the microscopic level densities from Hilaire's table (MLDH) \cite{mldh}. We repeated our calculations with these different level density model options, as shown in Fig. \ref{figt1}. Again, the cross sections are similar at lower energies regardless of the level density model used, but at high bombarding energies the results start to deviate. We also used the obtained cross sections to compute the reaction rates using Eq. (\ref{eqn3}) and the rates are similar in magnitude for the different level density models at astrophysically relevant temperature region.

%and the results are shown in Fig. \ref{figt2}. Within the astrophysically relevant temperature region, the results are similar for the different level density models. 

%\begin{figure}[!htb]
%\begin{center}
%\includegraphics[angle=270, totalheight=0.35\textheight, width=0.5\textwidth]{graph_trld.ps}
%\caption{ (Color online): The reaction rates for $^{26}$Al$(n,p)^{26}$Mg and $^{26}$Al$(n,\alpha)^{23}$Na calculated using the TALYS code with different level density model options.}
%\label{figt2}
%\end{center}
%\end{figure}
 
Another important input into the EMPIRE and TALYS codes are the optical model potentials.  EMPIRE incorporates the RIPL-2 \cite{ripl2} optical model potential library, which contains over 400 sets of parameters for nuclei up to Lr ($Z$ = 103) and energies up to 400 MeV. However, the default option for the optical model potential used for the reactions we studied are from Avrigeanu \textit{et al.} \cite{avrigeanu} for $\alpha$-particles and Koning and Delaroche \cite{koning} for protons and neutrons. Other optical model potentials that were used in our work are from Ferrer \textit{et al.} \cite{ferrer}, Harper and Alford \cite{harper} and Yamamuro \textit{et al.} \cite{yamamuro} for neutrons; Menet \textit{et al.} \cite{menet} and Harper and Alford \cite{harper} for protons; and McFadden and Satchler \cite{mcfadden} and Huizenga \textit{et al.} \cite{huizenga} for $\alpha$-particles. In contrast, TALYS employs the local and global parameterization of Koning and Delaroche \cite{koning} for the default neutron and proton optical model potentials, while for complex particles like deuterons, tritons, $^{3}$He and $\alpha$-particles the folding potential approach according to Watanabe \cite{watanabe} is used. Other options are the semi-microscopic optical model based on the Br$\ddot{u}$ckner-Hartree-Fock work of Jeukenne, Lejeune and Mahaux (JLM) \cite{jlm1, jlm2, jlm3, jlm4}. The optical models from McFadden and Satchler \cite{mcfadden} are also available for $\alpha$-particles. All of the optical model potentials mentioned so far are explored in this study.

\begin{figure}[!htb]
\begin{center}
\includegraphics[angle=270, totalheight=0.43\textheight, width=0.5\textwidth]{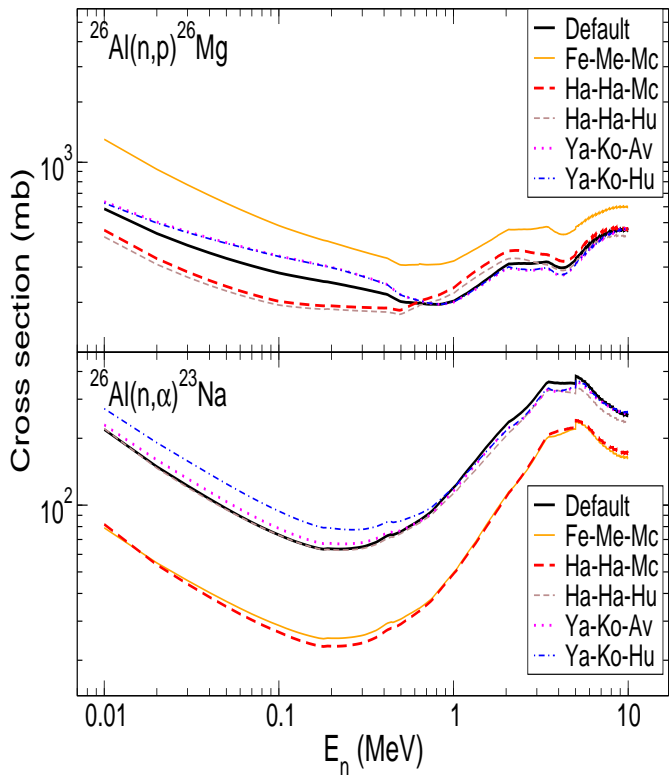}
\caption{ (Color online):  The $^{26}$Al($n,p)^{26}$Mg and $^{26}$Al$(n,\alpha)^{23}$Na cross sections as calculated using the EMPIRE code using different optical model potentials. In the legends, the first two letters denote the optical  model parameters used for neutrons, the next two denote those for protons, while the last two denote those for $\alpha$-particles; ``Fe'' stands for optical models from Ferrer \textit{et al.} \cite{ferrer}, ``Me'' stands for Menet \textit{et al.} \cite{menet}, ``Mc'' for McFadden and Satchler \cite{mcfadden}, ``Ha'' for Harper and Alford \cite{harper}, ``Hu'' for Huizenga \textit{et al.} \cite{huizenga}, ``Ya'' for Yamamuro \textit{et al.} \cite{yamamuro}, ``Ko'' for Koning and Delaroche \cite{koning}, and ``Av'' for Avrigeanu  \textit{et al.} \cite{avrigeanu}. The default option is described in the text. }
\label{fig3}
\end{center}
\end{figure}

\begin{figure}[!htb]
\begin{center}
\includegraphics[angle=270, totalheight=0.43\textheight, width=0.5\textwidth]{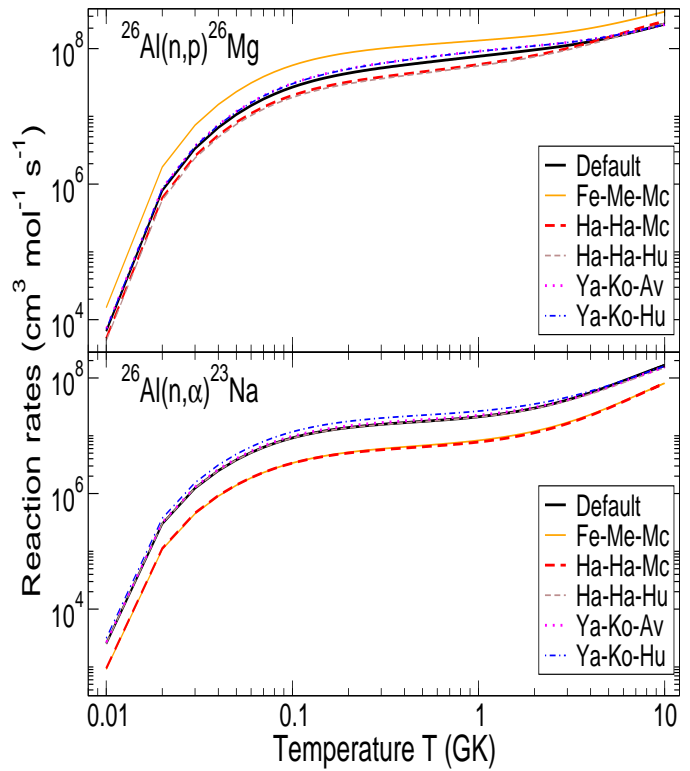}
\caption{ (Color online): The reaction rates for $^{26}$Al$(n,p)^{26}$Mg and $^{26}$Al$(n,\alpha)^{23}$Na calculated using the EMPIRE code with different optical model potentials. The legends are the same as in Fig. \ref{fig3}.}
\label{fig4}
\end{center}
\end{figure}

\begin{figure}[!htb]
\begin{center}
\includegraphics[angle=270, totalheight=0.43\textheight, width=0.5\textwidth]{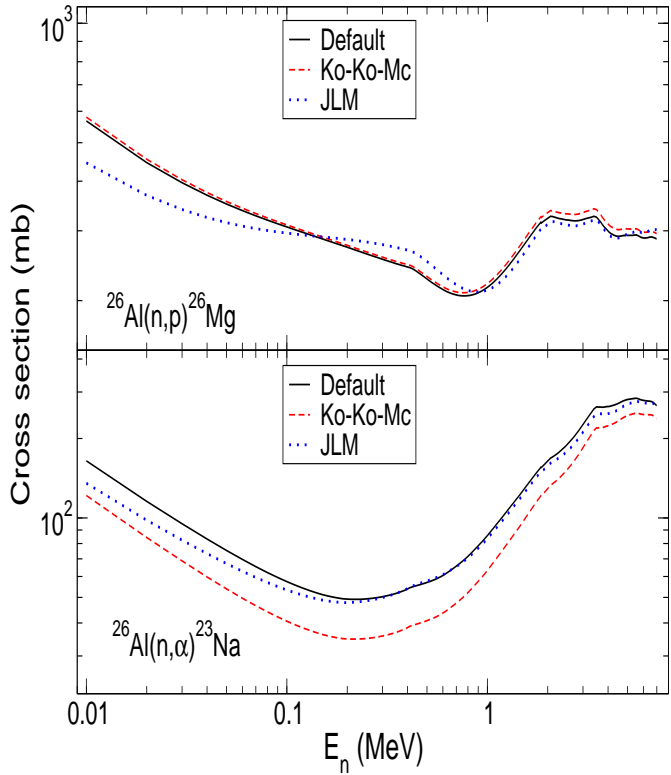}
\caption{ (Color online) The $^{26}$Al$(n,p)^{26}$Mg and $^{26}$Al$(n,\alpha)^{23}$Na cross sections as calculated using the TALYS code with different optical model potentials. The red-dashed line (``Ko-Ko-Mc'') represents the optical model parameter set where Koning and Delaroche \cite{koning} was used for neutrons and protons, while McFadden and Satchler \cite{mcfadden} was used for $\alpha$-particles; the blue-dotted line  (``JLM'') represents semi-microscopic optical models \cite{jlm1, jlm2, jlm3, jlm4}. The default option is described in the text.}
\label{fig5}
\end{center}
\end{figure}

\begin{figure}[!htb]
\begin{center}
\includegraphics[angle=270, totalheight=0.43\textheight, width=0.5\textwidth]{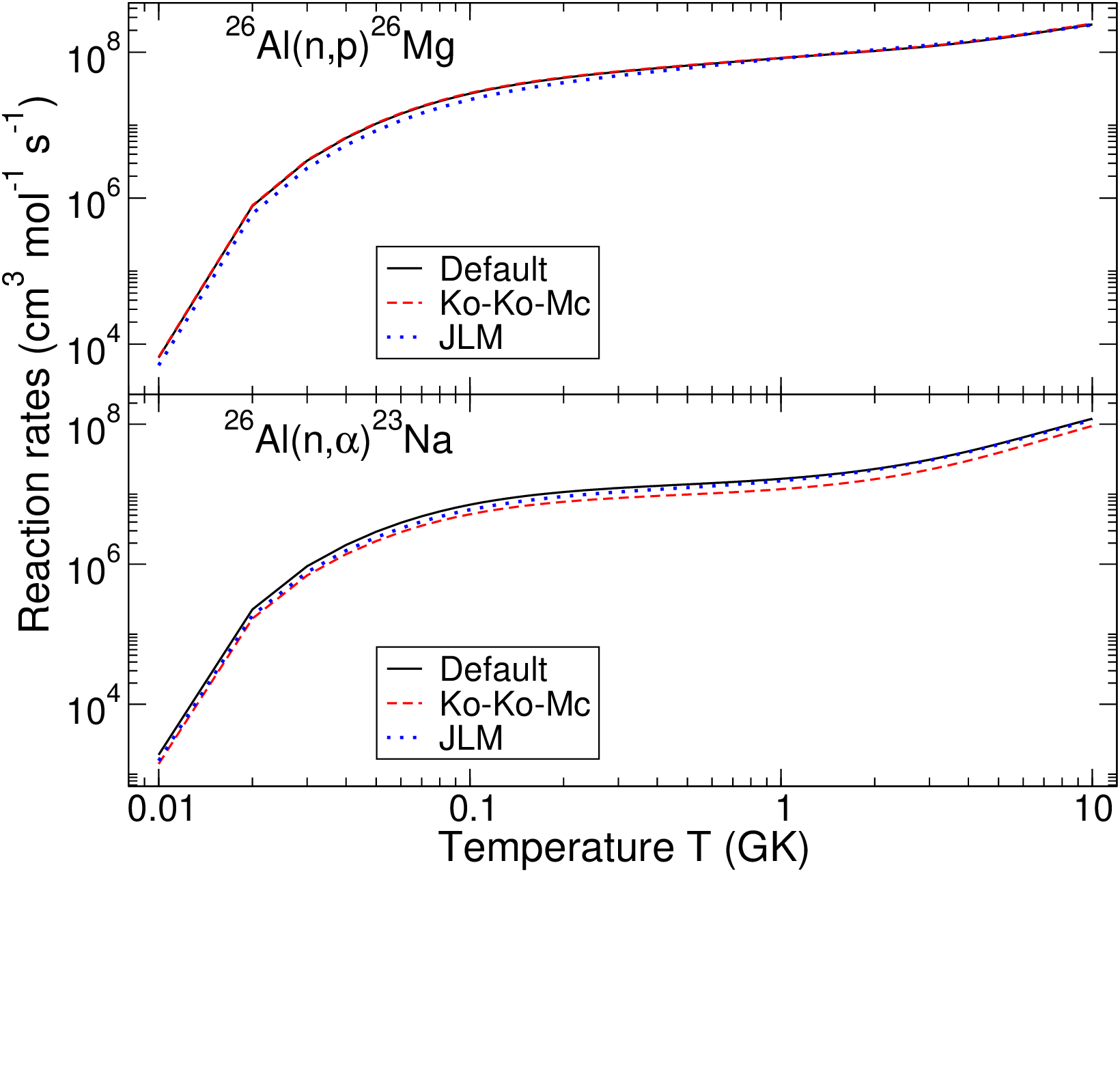}
\caption{ (Color online) Reaction rates for $^{26}$Al$(n,p)^{26}$Mg and $^{26}$Al$(n,\alpha)^{23}$Na as calculated using the TALYS code with different optical model potential sets; the legends are the same as in Fig. \ref{fig5}.}
\label{fig6}
\end{center}
\end{figure}

Unlike the situation for different level density models, the choice of optical model potential led to much larger differences in both the cross sections and the reaction rates. Figures \ref{fig3} and \ref{fig4} show the cross sections and reaction rates, respectively, for the different optical model potentials used in the EMPIRE calculations. The different prescriptions give rise to rates differences up to a factor of 3 for the $^{26}$Al$(n,p)^{26}$Mg reaction, while for $^{26}$Al$(n,\alpha)^{23}$Na the changes amount to a factor of 3.5. For the TALYS calculations, the cross sections and reaction rates are shown in Figs. \ref{fig5} and \ref{fig6}, respectively. The ratio of maximum to minimum rates for the three sets of optical model potentials used for the two reactions were within a factor of 2. These results demonstrate that the reaction rate does depend sensitively on the optical model potentials used to calculate the cross sections.
%\clearpage

\section{REACTION RATES} \label{sec4}

Using the experimental data reported in the literature and the theoretical 
calculations presented here, we estimate new rates for the reactions $^{26}$Al$(n,p)^{26}$Mg and $^{26}$Al$(n,\alpha)^{23}$Na. Since for these two reactions the available experimental information 
is rather sparse, it should be obvious to the reader that the rates cannot be 
evaluated in a statistically meaningful way, as was done, for example, in Ref. 
\cite{longilia}. Our modest goal at this stage is to 
provide updated reaction rates and to estimate reasonable uncertainties. 
Needless to say that our evaluated rates for these two reactions are no
substitute for future measurements. 

Our strategy was as follows. At low temperatures (T $<$ 100 MK) one cannot 
expect the theoretical rates to be reliable, since the level densities involved 
are rather small. Thus we adopted here the experimental rates, such that 
our low and high rate encompasses the range of uncertainty of the 
experimental rates. At higher temperatures we adopted the rates from our 
calculations (and from the experiment of Trautvetter \textit {et al.} \cite{traut} for the $^{26}$Al$(n,p)^{26}$Mg reaction).
To be more specific, our low and high rates corresponded to the extreme low and high values that resulted from our variation of reaction model parameters. The results of these parameter variations are displayed in Figs. \ref{fig4} and \ref{fig6}. In the intermediate temperature regime, near T = 0.1 GK, the low and high rates at small and large temperatures are matched smoothly using polynomial fits.  Finally, we estimate a recommended rate 
from the geometric mean value of the low and high rate at each temperature.

\begin{figure}[!htb]
\begin{center}
\includegraphics[angle=270, totalheight=0.43\textheight, width=0.5\textwidth]{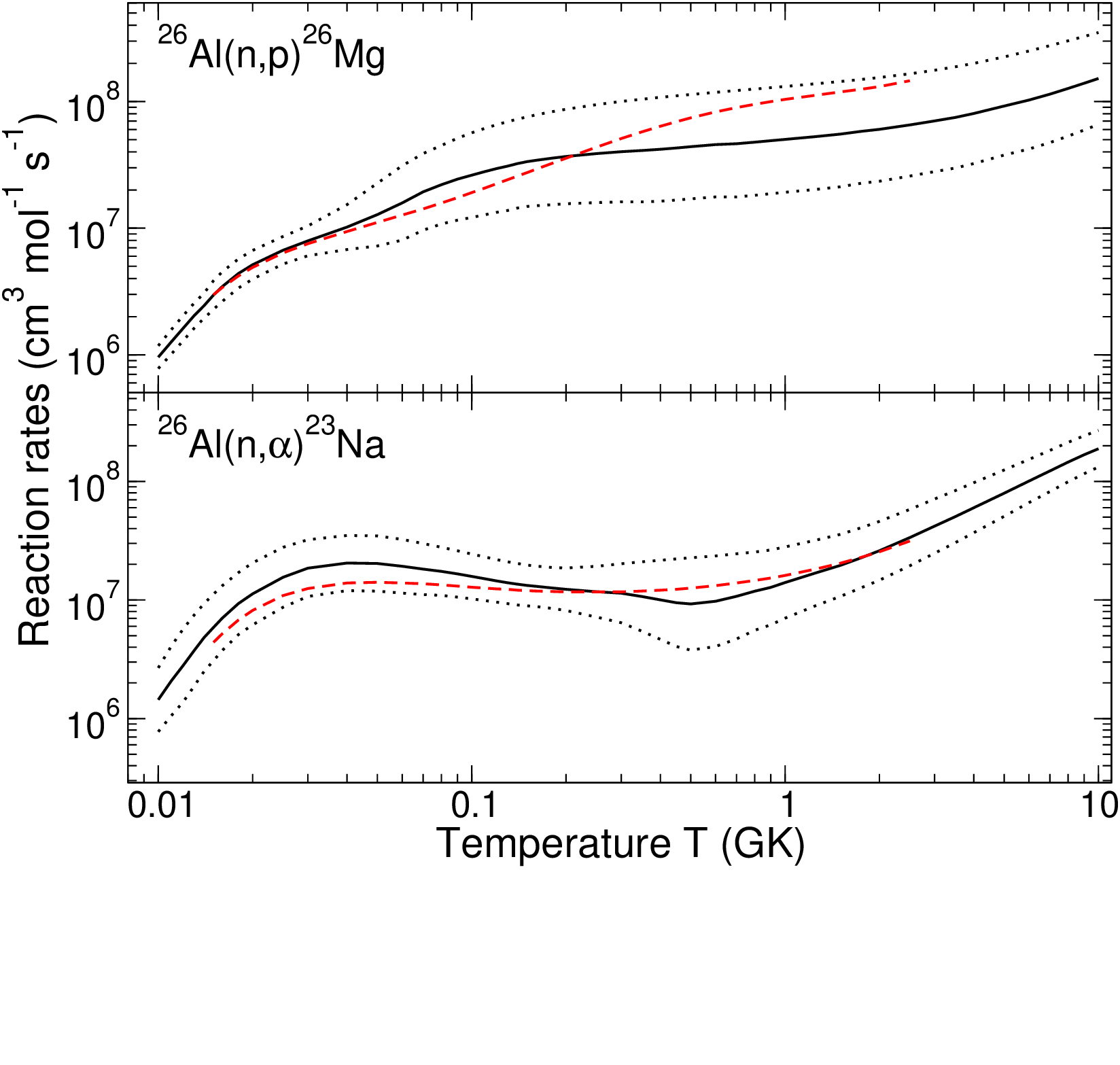}
\caption{ (Color online) Recommended rates obtained from experimental data and theoretical calculations. The dotted lines indicate the high and low rates, the black line represents their geometric mean, while the red-dashed line corresponds to the rate from Iliadis \textit {et al.} \cite{ilia}. All rates shown here account for thermal target excitations.}
\label{figreco21}
\end{center}
\end{figure}

The present study was motivated by a recent large scale sensitivity study of 
$^{26}$Al synthesis in massive stars (Iliadis \textit{et al.} \cite{ilia}). The rates for the 
$^{26}$Al$(n,p)^{26}$Mg and $^{26}$Al$(n,\alpha)^{23}$Na reactions used in that work were similar, but not identical, to the 
present recommened rates. No attempt was made in Ref. \cite{ilia} to estimate 
the actual range of rate uncertainty and, thus, the present results supersede
those of the former work. Figure \ref{figreco21} shows the present rates  and how they compare with the rates from Ref. \cite{ilia}.  The low, recommended, and high stellar rates are given in Tables \ref{tab1} and \ref{tab2}, where the values listed account for thermal target excitations.

 %These data are at low temperature while the theoretical calculations are extended to 10 GK. The upper and lower limits on the rates were determined such that they enclose the experimental data at low temperatures and the theoretical calculations at high temperatures while also accounting for the systematic uncertainty in the calculated rates as a result of the optical model parameters. We used polynomial fits at the intersections of the experimental and calculated rate to ensure a smooth transition. The median rates were obtained from the geometric mean of the upper and lower limits. 

\begin{table}[!hbt]
\begin{center}
\caption{Recommended stellar rates for $^{26}$Al($n,p)^{26}$Mg in units of cm$^{3}$mol$^{-1}$s$^{-1}$.}
\begin{tabular}{cccc}
\hline
\hline
   T (GK)    &       low rate       &       median rate       &         high rate      \\
%{} & (cm$^{3}$mol$^{-1}$s$^{-1}$)  & (cm$^{3}$mol$^{-1}$s$^{-1}$)   & (cm$^{3}$mol$^{-1}$s$^{-1}$)\\
\hline
{0.010} & {7.79$\times 10^{5}$} & {9.58$\times 10^{5}$} & {1.18$\times 10^{6}$}\\
{0.011} & {1.01$\times 10^{6}$} & {1.26$\times 10^{6}$} & {1.58$\times 10^{6}$}\\
{0.012} & {1.29$\times 10^{6}$} & {1.62$\times 10^{6}$} & {2.04$\times 10^{6}$}\\
{0.013} & {1.62$\times 10^{6}$} & {2.03$\times 10^{6}$} & {2.56$\times 10^{6}$}\\
{0.014} & {1.94$\times 10^{6}$} & {2.44$\times 10^{6}$} & {3.07$\times 10^{6}$}\\
{0.015} & {2.28$\times 10^{6}$} & {2.96$\times 10^{6}$} & {3.84$\times 10^{6}$}\\
{0.016} & {2.64$\times 10^{6}$} & {3.46$\times 10^{6}$} & {4.53$\times 10^{6}$}\\
{0.018} & {3.36$\times 10^{6}$} & {4.38$\times 10^{6}$} & {5.71$\times 10^{6}$}\\
{0.020} & {3.96$\times 10^{6}$} & {5.14$\times 10^{6}$} & {6.68$\times 10^{6}$}\\
{0.025} & {5.21$\times 10^{6}$} & {6.69$\times 10^{6}$} & {8.60$\times 10^{6}$}\\
{0.030} & {6.05$\times 10^{6}$} & {7.94$\times 10^{6}$} & {1.04$\times 10^{7}$}\\
{0.040} & {6.78$\times 10^{6}$} & {1.02$\times 10^{7}$} & {1.53$\times 10^{7}$}\\
{0.050} & {7.23$\times 10^{6}$} & {1.28$\times 10^{7}$} & {2.26$\times 10^{7}$}\\
{0.060} & {8.05$\times 10^{6}$} & {1.58$\times 10^{7}$} & {3.11$\times 10^{7}$}\\
{0.070} & {9.63$\times 10^{6}$} & {1.93$\times 10^{7}$} & {3.88$\times 10^{7}$}\\
{0.080} & {1.08$\times 10^{7}$} & {2.20$\times 10^{7}$} & {4.50$\times 10^{7}$}\\
{0.090} & {1.16$\times 10^{7}$} & {2.43$\times 10^{7}$} & {5.13$\times 10^{7}$}\\
{0.100} & {1.21$\times 10^{7}$} & {2.62$\times 10^{7}$} & {5.66$\times 10^{7}$}\\
{0.110} & {1.27$\times 10^{7}$} & {2.79$\times 10^{7}$} & {6.13$\times 10^{7}$}\\
{0.120} & {1.33$\times 10^{7}$} & {2.95$\times 10^{7}$} & {6.55$\times 10^{7}$}\\
{0.130} & {1.38$\times 10^{7}$} & {3.09$\times 10^{7}$} & {6.92$\times 10^{7}$}\\
{0.140} & {1.44$\times 10^{7}$} & {3.23$\times 10^{7}$} & {7.26$\times 10^{7}$}\\
{0.150} & {1.48$\times 10^{7}$} & {3.35$\times 10^{7}$} & {7.56$\times 10^{7}$}\\
{0.160} & {1.50$\times 10^{7}$} & {3.43$\times 10^{7}$} & {7.83$\times 10^{7}$}\\
{0.180} & {1.53$\times 10^{7}$} & {3.57$\times 10^{7}$} & {8.30$\times 10^{7}$}\\
{0.200} & {1.56$\times 10^{7}$} & {3.68$\times 10^{7}$} & {8.69$\times 10^{7}$}\\
{0.250} & {1.59$\times 10^{7}$} & {3.88$\times 10^{7}$} & {9.45$\times 10^{7}$}\\
{0.300} & {1.61$\times 10^{7}$} & {4.02$\times 10^{7}$} & {1.00$\times 10^{8}$}\\
{0.350} & {1.61$\times 10^{7}$} & {4.10$\times 10^{7}$} & {1.04$\times 10^{8}$}\\
{0.400} & {1.63$\times 10^{7}$} & {4.20$\times 10^{7}$} & {1.08$\times 10^{8}$}\\
{0.450} & {1.67$\times 10^{7}$} & {4.30$\times 10^{7}$} & {1.11$\times 10^{8}$}\\
{0.500} & {1.70$\times 10^{7}$} & {4.40$\times 10^{7}$} & {1.14$\times 10^{8}$}\\
{0.600} & {1.77$\times 10^{7}$} & {4.57$\times 10^{7}$} & {1.18$\times 10^{8}$}\\
{0.700} & {1.77$\times 10^{7}$} & {4.64$\times 10^{7}$} & {1.22$\times 10^{8}$}\\
{0.800} & {1.82$\times 10^{7}$} & {4.78$\times 10^{7}$} & {1.25$\times 10^{8}$}\\
{0.900} & {1.87$\times 10^{7}$} & {4.90$\times 10^{7}$} & {1.29$\times 10^{8}$}\\
{1.000} & {1.92$\times 10^{7}$} & {5.02$\times 10^{7}$} & {1.31$\times 10^{8}$}\\
{1.250} & {2.02$\times 10^{7}$} & {5.28$\times 10^{7}$} & {1.38$\times 10^{8}$}\\
{1.500} & {2.12$\times 10^{7}$} & {5.53$\times 10^{7}$} & {1.44$\times 10^{8}$}\\
{1.750} & {2.26$\times 10^{7}$} & {5.81$\times 10^{7}$} & {1.49$\times 10^{8}$}\\
{2.000} & {2.35$\times 10^{7}$} & {6.02$\times 10^{7}$} & {1.55$\times 10^{8}$}\\
{2.500} & {2.58$\times 10^{7}$} & {6.53$\times 10^{7}$} & {1.65$\times 10^{8}$}\\
{3.000} & {2.79$\times 10^{7}$} & {7.02$\times 10^{7}$} & {1.76$\times 10^{8}$}\\
{3.500} & {2.99$\times 10^{7}$} & {7.49$\times 10^{7}$} & {1.88$\times 10^{8}$}\\
{4.000} & {3.24$\times 10^{7}$} & {8.04$\times 10^{7}$} & {2.00$\times 10^{8}$}\\
{5.000} & {3.78$\times 10^{7}$} & {9.22$\times 10^{7}$} & {2.25$\times 10^{8}$}\\
{6.000} & {4.21$\times 10^{7}$} & {1.03$\times 10^{8}$} & {2.51$\times 10^{8}$}\\
{7.000} & {4.73$\times 10^{7}$} & {1.14$\times 10^{8}$} & {2.76$\times 10^{8}$}\\
{8.000} & {5.33$\times 10^{7}$} & {1.27$\times 10^{8}$} & {3.02$\times 10^{8}$}\\
{9.000} & {5.96$\times 10^{7}$} & {1.39$\times 10^{8}$} & {3.26$\times 10^{8}$}\\
{       10.00              }  &  {              6.59 $\times 10^{7}$              }  &  {              1.52 $\times 10^{8}$              }  &  {              3.50 $\times 10^{8}$              } \\
\hline
\hline
\end{tabular}

\label{tab1}

\end{center}
\end{table}

\begin{table}[!hbt]
\begin{center}
\caption{Recommended stellar rates for $^{26}$Al($n,\alpha$)$^{23}$Na in units of cm$^{3}$mol$^{-1}$s$^{-1}$.}
\begin{tabular}{cccc}
\hline
\hline
T (GK) &low rate&median rate&  high rate \\
\hline
{0.010} & {7.76$\times 10^{5}$} & {1.44$\times 10^{6}$} & {2.68$\times 10^{6}$}\\
{0.011} & {1.06$\times 10^{6}$} & {2.08$\times 10^{6}$} & {4.08$\times 10^{6}$}\\
{0.012} & {1.38$\times 10^{6}$} & {2.80$\times 10^{6}$} & {5.69$\times 10^{6}$}\\
{0.013} & {1.87$\times 10^{6}$} & {3.74$\times 10^{6}$} & {7.45$\times 10^{6}$}\\
{0.014} & {2.50$\times 10^{6}$} & {4.83$\times 10^{6}$} & {9.32$\times 10^{6}$}\\
{0.015} & {3.07$\times 10^{6}$} & {5.87$\times 10^{6}$} & {1.12$\times 10^{7}$}\\
{0.016} & {3.75$\times 10^{6}$} & {7.04$\times 10^{6}$} & {1.32$\times 10^{7}$}\\
{0.018} & {5.10$\times 10^{6}$} & {9.31$\times 10^{6}$} & {1.70$\times 10^{7}$}\\
{0.020} & {6.17$\times 10^{6}$} & {1.13$\times 10^{7}$} & {2.06$\times 10^{7}$}\\
{0.025} & {8.68$\times 10^{6}$} & {1.55$\times 10^{7}$} & {2.78$\times 10^{7}$}\\
{0.030} & {1.07$\times 10^{7}$} & {1.86$\times 10^{7}$} & {3.22$\times 10^{7}$}\\
{0.040} & {1.20$\times 10^{7}$} & {2.05$\times 10^{7}$} & {3.50$\times 10^{7}$}\\
{0.050} & {1.19$\times 10^{7}$} & {2.03$\times 10^{7}$} & {3.47$\times 10^{7}$}\\
{0.060} & {1.14$\times 10^{7}$} & {1.92$\times 10^{7}$} & {3.24$\times 10^{7}$}\\
{0.070} & {1.11$\times 10^{7}$} & {1.82$\times 10^{7}$} & {2.99$\times 10^{7}$}\\
{0.080} & {1.09$\times 10^{7}$} & {1.74$\times 10^{7}$} & {2.78$\times 10^{7}$}\\
{0.090} & {1.06$\times 10^{7}$} & {1.66$\times 10^{7}$} & {2.60$\times 10^{7}$}\\
{0.100} & {1.02$\times 10^{7}$} & {1.58$\times 10^{7}$} & {2.44$\times 10^{7}$}\\
{0.110} & {9.89$\times 10^{6}$} & {1.51$\times 10^{7}$} & {2.31$\times 10^{7}$}\\
{0.120} & {9.58$\times 10^{6}$} & {1.45$\times 10^{7}$} & {2.19$\times 10^{7}$}\\
{0.130} & {9.32$\times 10^{6}$} & {1.40$\times 10^{7}$} & {2.10$\times 10^{7}$}\\
{0.140} & {9.08$\times 10^{6}$} & {1.36$\times 10^{7}$} & {2.03$\times 10^{7}$}\\
{0.150} & {8.93$\times 10^{6}$} & {1.33$\times 10^{7}$} & {1.97$\times 10^{7}$}\\
{0.160} & {8.81$\times 10^{6}$} & {1.30$\times 10^{7}$} & {1.93$\times 10^{7}$}\\
{0.180} & {8.50$\times 10^{6}$} & {1.26$\times 10^{7}$} & {1.88$\times 10^{7}$}\\
{0.200} & {8.14$\times 10^{6}$} & {1.23$\times 10^{7}$} & {1.86$\times 10^{7}$}\\
{0.250} & {7.18$\times 10^{6}$} & {1.17$\times 10^{7}$} & {1.92$\times 10^{7}$}\\
{0.300} & {6.43$\times 10^{6}$} & {1.14$\times 10^{7}$} & {2.02$\times 10^{7}$}\\
{0.350} & {5.44$\times 10^{6}$} & {1.07$\times 10^{7}$} & {2.10$\times 10^{7}$}\\
{0.400} & {4.65$\times 10^{6}$} & {1.00$\times 10^{7}$} & {2.16$\times 10^{7}$}\\
{0.450} & {4.05$\times 10^{6}$} & {9.48$\times 10^{6}$} & {2.22$\times 10^{7}$}\\
{0.500} & {3.77$\times 10^{6}$} & {9.26$\times 10^{6}$} & {2.27$\times 10^{7}$}\\
{0.600} & {4.05$\times 10^{6}$} & {9.76$\times 10^{6}$} & {2.35$\times 10^{7}$}\\
{0.700} & {4.69$\times 10^{6}$} & {1.07$\times 10^{7}$} & {2.45$\times 10^{7}$}\\
{0.800} & {5.55$\times 10^{6}$} & {1.19$\times 10^{7}$} & {2.54$\times 10^{7}$}\\
{0.900} & {6.17$\times 10^{6}$} & {1.28$\times 10^{7}$} & {2.64$\times 10^{7}$}\\
{1.000} & {7.02$\times 10^{6}$} & {1.40$\times 10^{7}$} & {2.80$\times 10^{7}$}\\
{1.250} & {8.94$\times 10^{6}$} & {1.69$\times 10^{7}$} & {3.19$\times 10^{7}$}\\
{1.500} & {1.07$\times 10^{7}$} & {1.96$\times 10^{7}$} & {3.57$\times 10^{7}$}\\
{1.750} & {1.27$\times 10^{7}$} & {2.27$\times 10^{7}$} & {4.07$\times 10^{7}$}\\
{2.000} & {1.47$\times 10^{7}$} & {2.60$\times 10^{7}$} & {4.60$\times 10^{7}$}\\
{2.500} & {1.94$\times 10^{7}$} & {3.36$\times 10^{7}$} & {5.81$\times 10^{7}$}\\
{3.000} & {2.48$\times 10^{7}$} & {4.20$\times 10^{7}$} & {7.10$\times 10^{7}$}\\
{3.500} & {3.05$\times 10^{7}$} & {5.06$\times 10^{7}$} & {8.38$\times 10^{7}$}\\
{4.000} & {3.69$\times 10^{7}$} & {6.00$\times 10^{7}$} & {9.74$\times 10^{7}$}\\
{5.000} & {5.07$\times 10^{7}$} & {7.95$\times 10^{7}$} & {1.25$\times 10^{8}$}\\
{6.000} & {6.61$\times 10^{7}$} & {1.01$\times 10^{8}$} & {1.53$\times 10^{8}$}\\
{7.000} & {8.21$\times 10^{7}$} & {1.22$\times 10^{8}$} & {1.83$\times 10^{8}$}\\
{8.000} & {9.93$\times 10^{7}$} & {1.46$\times 10^{8}$} & {2.13$\times 10^{8}$}\\
{9.000} & {1.16$\times 10^{8}$} & {1.68$\times 10^{8}$} & {2.42$\times 10^{8}$}\\
{       10.00              }  &  {              1.32 $\times 10^{8}$              }  &  {              1.89 $\times 10^{8}$              }  &  {              2.69 $\times 10^{8}$              } \\
\hline
\hline
\end{tabular}

\label{tab2}

\end{center}
\end{table}

%\clearpage

\section{CONCLUSION}
In this paper, we compared cross sections and reaction rates for $^{26}$Al$(n,p)^{26}$Mg and $^{26}$Al$(n,\alpha)^{23}$Na obtained from the EMPIRE, TALYS and NON-SMOKER nuclear reaction codes. We have also shown how the changes in the level density models and optical model potentials affect the calculations.  The reaction rates derived from EMPIRE, TALYS and NON-SMOKER are generally consistent within a factor of 2 for most astrophysically relevant temperatures. In particular, the reaction rates obtained using  EMPIRE and TALYS agree within a factor of 1.3 at all temperatures, assuming the same input parameters for each code. We have also presented recommended rates, which are based on the available data and the results from the theoretical calculations.

\appendix

\section{Reaction rates for the isomeric target state of $^{26}$Al}

\begin{figure}[!htb]
\begin{center}
\includegraphics[angle=270, totalheight=0.35\textheight, width=0.5\textwidth]{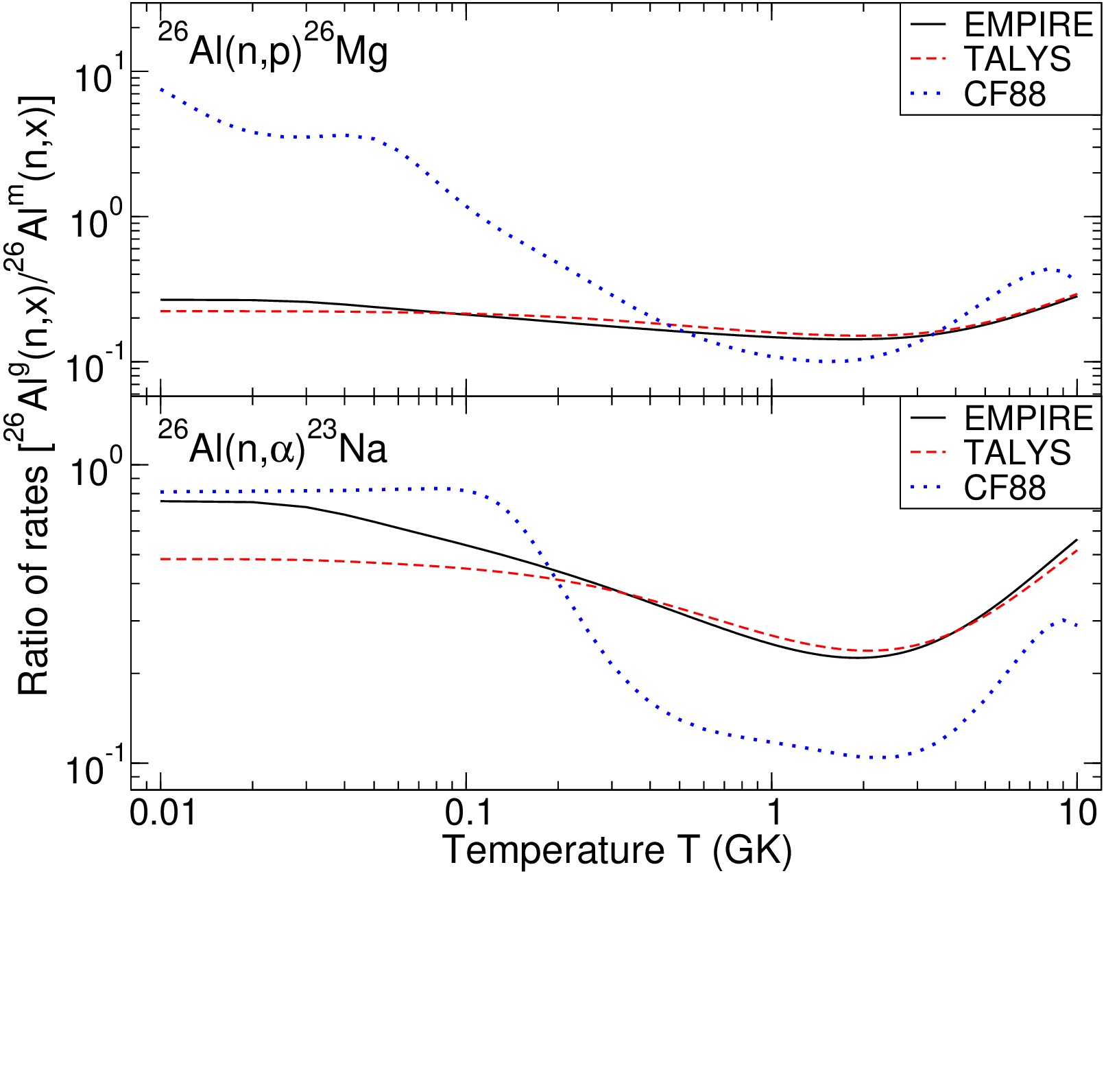}
\caption{ (Color online): Ratio of reaction rates for the ground state target with respect to the isomeric state target. The results from codes EMPIRE \cite{empire} and TALYS \cite{talys} are compared with the CF88 compilation \cite{caughlan}.}
\label{gm1}
\end{center}
\end{figure}

EMPIRE and TALYS have the option to perform calculations assuming that the target is in an excited state. This is a very useful feature since in some situations the rates for the ground and isomeric states in $^{26}$Al need to be known separately. In Fig. \ref{gm1}, we show the ratio of the reaction rates for $^{26}$Al in its ground state and its isomeric state. The only tabulated rates involving the isomeric state, $^{26}$Al$^{m}$, as the target are from Caughlan and Fowler (CF88) \cite{caughlan}. It is not clear how the CF88 rates were calculated and their results clearly disagree with our calculations, both in magnitude and temperature dependence. The results of EMPIRE and TALYS are consistent to within a factor of 1.2 for the $^{26}$Al$^{m}(n,p)^{26}$Mg reaction and within a factor of 1.6 for the $^{26}$Al$^{m}(n,\alpha)^{23}$Na reaction. Table \ref{tab3}  lists the reaction rates for $^{26}$Al$^{m}(n,p)^{26}$Mg and $^{26}$Al$^{m}(n,\alpha)^{23}$Na. The rates represent the mean values calculated from the EMPIRE and TALYS codes at each temperature.

%Tables \ref{tab3} and \ref{tab4} show the reaction rates calculated with the EMPIRE and TALYS codes for $^{26}$Al$_{m}(n,p)^{26}$Mg and $^{26}$Al$_{m}(n,\alpha)^{23}$Na respectively, where $^{26}$Al$_{m}$ represents the isomeric target.

%\clearpage

\begin{table}[!hbt]
\begin{center}
\caption{Rates for the $^{26}$Al$^{m}(n,p)^{26}$Mg and $^{26}$Al$^{m}(n,\alpha)^{23}$Na reactions as calculated using EMPIRE and TALYS codes. The rates are evaluated from the mean values of the two codes and are in units of cm$^{3}$mol$^{-1}$s$^{-1}$.}
\begin{tabular}{ccc}
\hline
\hline
{T (GK)       } & {               $^{26}$Al$^{m}(n,p)^{26}$Mg            }  & {                            $^{26}$Al$^{m}(n,\alpha)^{23}$Na} \\
{} & rates & rates \\
\hline
{0.010} & {2.73 $\times 10^{4}$} & {3.63 $\times 10^{3}$}\\
{0.011} & {6.79 $\times 10^{4}$} & {9.05 $\times 10^{3}$}\\
{0.012} & {1.44 $\times 10^{5}$} & {1.92 $\times 10^{4}$}\\
{0.013} & {2.68 $\times 10^{5}$} & {3.57 $\times 10^{4}$}\\
{0.014} & {4.54 $\times 10^{5}$} & {6.05 $\times 10^{4}$}\\
{0.015} & {7.12 $\times 10^{5}$} & {9.47 $\times 10^{4}$}\\
{0.016} & {1.04 $\times 10^{6}$} & {1.40 $\times 10^{5}$}\\
{0.018} & {1.99 $\times 10^{6}$} & {2.64 $\times 10^{5}$}\\
{0.020} & {3.26 $\times 10^{6}$} & {4.34 $\times 10^{5}$}\\
{0.025} & {7.79 $\times 10^{6}$} & {1.04 $\times 10^{6}$}\\
{0.030} & {1.39 $\times 10^{7}$} & {1.84 $\times 10^{6}$}\\
{0.040} & {2.89 $\times 10^{7}$} & {3.82 $\times 10^{6}$}\\
{0.050} & {4.61 $\times 10^{7}$} & {6.07 $\times 10^{6}$}\\
{0.060} & {6.40 $\times 10^{7}$} & {8.37 $\times 10^{6}$}\\
{0.070} & {8.13 $\times 10^{7}$} & {1.06 $\times 10^{7}$}\\
{0.080} & {9.76 $\times 10^{7}$} & {1.27 $\times 10^{7}$}\\
{0.090} & {1.13 $\times 10^{8}$} & {1.47 $\times 10^{7}$}\\
{0.100} & {1.27 $\times 10^{8}$} & {1.65 $\times 10^{7}$}\\
{0.110} & {1.40 $\times 10^{8}$} & {1.81 $\times 10^{7}$}\\
{0.120} & {1.52 $\times 10^{8}$} & {1.96 $\times 10^{7}$}\\
{0.130} & {1.64 $\times 10^{8}$} & {2.10 $\times 10^{7}$}\\
{0.140} & {1.74 $\times 10^{8}$} & {2.23 $\times 10^{7}$}\\
{0.150} & {1.84 $\times 10^{8}$} & {2.36 $\times 10^{7}$}\\
{0.160} & {1.93 $\times 10^{8}$} & {2.47 $\times 10^{7}$}\\
{0.180} & {2.10 $\times 10^{8}$} & {2.69 $\times 10^{7}$}\\
{0.200} & {2.25 $\times 10^{8}$} & {2.88 $\times 10^{7}$}\\
{0.250} & {2.59 $\times 10^{8}$} & {3.30 $\times 10^{7}$}\\
{0.300} & {2.87 $\times 10^{8}$} & {3.67 $\times 10^{7}$}\\
{0.350} & {3.12 $\times 10^{8}$} & {4.00 $\times 10^{7}$}\\
{0.400} & {3.35 $\times 10^{8}$} & {4.31 $\times 10^{7}$}\\
{0.450} & {3.56 $\times 10^{8}$} & {4.60 $\times 10^{7}$}\\
{0.500} & {3.75 $\times 10^{8}$} & {4.88 $\times 10^{7}$}\\
{0.600} & {4.11 $\times 10^{8}$} & {5.41 $\times 10^{7}$}\\
{0.700} & {4.42 $\times 10^{8}$} & {5.92 $\times 10^{7}$}\\
{0.800} & {4.71 $\times 10^{8}$} & {6.41 $\times 10^{7}$}\\
{0.900} & {4.96 $\times 10^{8}$} & {6.88 $\times 10^{7}$}\\
{1.000} & {5.20 $\times 10^{8}$} & {7.34 $\times 10^{7}$}\\
{1.250} & {5.72 $\times 10^{8}$} & {8.46 $\times 10^{7}$}\\
{1.500} & {6.14 $\times 10^{8}$} & {9.54 $\times 10^{7}$}\\
{1.750} & {6.49 $\times 10^{8}$} & {1.05 $\times 10^{8}$}\\
{2.000} & {6.78 $\times 10^{8}$} & {1.15 $\times 10^{8}$}\\
{2.500} & {7.23 $\times 10^{8}$} & {1.33 $\times 10^{8}$}\\
{3.000} & {7.56 $\times 10^{8}$} & {1.49 $\times 10^{8}$}\\
{3.500} & {7.79 $\times 10^{8}$} & {1.63 $\times 10^{8}$}\\
{4.000} & {7.96 $\times 10^{8}$} & {1.75 $\times 10^{8}$}\\
{5.000} & {8.17 $\times 10^{8}$} & {1.97 $\times 10^{8}$}\\
{6.000} & {8.27 $\times 10^{8}$} & {2.14 $\times 10^{8}$}\\
{7.000} & {8.32 $\times 10^{8}$} & {2.29 $\times 10^{8}$}\\
{8.000} & {8.34 $\times 10^{8}$} & {2.42 $\times 10^{8}$}\\
{9.000} & {8.34 $\times 10^{8}$} & {2.52 $\times 10^{8}$}\\
{       10.00              }  &  {              8.34 $\times 10^{8}$              }  &  {              2.62 $\times 10^{8}$              }   \\
\hline
\hline
\end{tabular}

\label{tab3}

\end{center}
\end{table}

%\clearpage

%\end{Large}

\end{document}